\documentstyle[preprint,prl,aps]{revtex}
\begin{document}
%\draft
%%%%%End of Preamble
%%%%Start of Text%%%%%%%%%%%%%%%%%%%%%%%%%%%%%%%%%%%%%%%%%%%%%%%%%%%%%%%
\preprint{
\vbox{\halign{&##\hfil\cr
        & cond-mat/9808088       \cr
        & August 1998            \cr
        & revised November 1998  \cr}}}

\title{Semiclassical Corrections to the Oscillation Frequencies
of a Trapped Bose-Einstein Condensate}

\author{Eric Braaten}
\address{Physics Department, Ohio State University, Columbus OH 43210}

\author{John Pearson}
\address{Department of Physics and Astronomy, University of Kentucky, 
Lexington KY 40506}

\maketitle

\begin{abstract}
The oscillation frequencies of 
collective excitations of a trapped Bose-Einstein condensate, 
when calculated in the mean-field approximation 
and in the Thomas-Fermi limit, are independent of the scattering length 
$a$. We calculate the leading corrections to the frequencies from quantum 
fluctuations around the mean field. The semiclassical correction 
is proportional to $N^{1/5} a^{6/5}$, 
where $N$ is the number of atoms in the condensate. 
The correction is positive semidefinite and is zero for surface modes 
whose eigenfunctions have a vanishing laplacian. 
The shift in the frequency of the lowest quadrupole mode 
for an axially symmetric trap is large enough that it should 
be measurable in future experiments.
\end{abstract}

\vfill \eject
\narrowtext

The achievement of Bose-Einstein condensation 
in trapped atomic vapors \cite{BEC} has opened up new possibilities 
for the experimental study of nonhomogeneous interacting Bose gases. 
This system can be conveniently described by a quantum field theory, but 
the quantum field equations are extremely difficult to solve in general. 
Fortunately, many of the basic properties of the condensates 
in existing experiments are described accurately by the 
mean-field approximation, which 
reduces the problem to solving the Gross-Pitaevskii 
equation, a classical equation for the mean field. 
These properties include the density profile of the ground state and 
collective oscillations of the condensate. 
Fluctuations of the quantum field around the mean field 
provide corrections to mean-field predictions that 
scale as the square root of the peak number density of the atoms. 
These corrections should be observable through precise 
measurements of the properties of condensates containing 
a large number of atoms.

Among the most beautiful experiments on Bose-Einstein condensates 
are the excitation and study of collective oscillations of the condensates. 
Among the properties of the collective modes that have been measured 
are the frequencies of the lowest normal modes at extremely low temperatures 
\cite{Jin-1,Mewes,Stamper-Kurn}, the temperature dependence of the 
frequencies and their damping rates, \cite{Stamper-Kurn,Jin-2}, 
and the speed of sound \cite{Andrews}. 
The frequencies of the normal modes can be measured very accurately. 
In one experiment, they have been measured 
to better than a fraction of a percent \cite{Stamper-Kurn}. 
These frequencies are therefore an ideal place to look for the effects of 
quantum field fluctuations.

Most previous studies of the collective excitations of a trapped condensate 
at zero temperature have been carried out using the mean-field 
approximation or the Bogoliubov approximation, which is equivalent at $T=0$. 
The collective excitations have been studied by 
variational methods \cite{variational}, 
by numerical methods \cite{numerical}, 
and by analytic methods \cite{Stringari,Ohberg-Fliesser}. 
In the Thomas-Fermi limit, 
the oscillation frequencies for a condensate in a 
harmonic potential can be obtained analytically and are independent of the 
interaction strength of the atoms and the number of atoms in the trap 
\cite{Stringari}. Quantum fluctuations around the mean field give 
fractional corrections to the frequencies that 
are proportional to $\sqrt{\rho a^3}$, where $\rho$ is the peak number density 
and $a$ is the S-wave scattering length of the atoms. 
The purpose of this paper is to calculate these corrections.

A Bose-Einstein condensate containing a large number $N$ 
of identical atoms in a trapping potential $V({\bf r})$ 
can be described by a quantum field theory with a single 
complex-valued field $\psi({\bf r},t)$ that satisfies 
\begin{equation} 
i \hbar \dot \psi \;=\; 
\left[ - {\hbar^2 \over 2 m} \nabla^2 + V - \mu \right] \psi 
\;+\; {\hbar^2 g \over 4 m} (\psi^\dagger \psi ) \psi, 
\label{qfe} 
\end{equation} 
where $g= 16 \pi a$. 
The chemical potential $\mu$ in (\ref{qfe}) is to be adjusted so that the 
expectation value of $\int d^3 r \psi^\dagger \psi$, 
the total number of atoms, is equal to $N$. 
A collective oscillation of the Bose-Einstein condensate is described 
by a time-dependent state in the quantum field theory. 
Letting $\langle \ldots \rangle$ denote the expectation value in that state, 
we can write the quantum field as $\psi = \phi + \tilde\psi$, 
where $\phi = \langle \psi \rangle$ is the mean field and 
$\tilde\psi$ is the quantum fluctuation field 
that satisfies $\langle \tilde\psi \rangle = 0$.
The number density $\rho$ and the current density ${\bf j}$ 
in that state are
\begin{eqnarray}
\rho &=& |\phi|^2 
\;+\; \langle \tilde\psi^{\dagger} \tilde\psi \rangle, 
\label{rho-SC}
\\
{\bf j} &=& -i {\hbar \over 2 m} \left( \phi^* \nabla \phi 
		- \phi \nabla \phi^* \right)
\;-\; i {\hbar \over 2 m} \langle \tilde\psi^{\dagger} \nabla \tilde\psi 
		- \nabla \tilde\psi^{\dagger} \tilde\psi \rangle.
\label{j-SC}
\end{eqnarray}
The continuity equation $\dot \rho + \nabla \cdot {\bf j} = 0$
follows from the U(1) symmetry of the field theory.
Taking the expectation value of (\ref{qfe}), 
we obtain a partial differential equation for $\phi$ 
that involves the matrix elements 
of $\tilde\psi$:
\begin{equation}
i \hbar \dot \phi
\;+\; \left[ -{\hbar^2 \over 2 m}\nabla^2 + V - \mu \right] \phi
\;+\; {\hbar^2 g \over 4m} \left[ |\phi|^2 \phi
	+ 2 \langle \tilde\psi^\dagger \tilde\psi \rangle \phi
	+ \langle \tilde\psi \tilde\psi \rangle \phi^*
	+ \langle \tilde\psi^\dagger \tilde\psi \tilde\psi \rangle
	\right].
\label{qfe-phi}
\end{equation}
In the mean-field approximation, the matrix elements are neglected 
and (\ref{qfe-phi}) reduces to the Gross-Pitaevskii equation. 
The effects of quantum field fluctuations are contained 
in the matrix elements, which can be expressed as functionals of the 
mean field $\phi$. They are dominated by modes of $\tilde\psi$ 
with wavelengths comparable to the local coherence length $\xi$,
which is the scale of the wavelength above which the
excitations of the condensate are modified by collective effects.
In the Thomas-Fermi limit, $\xi$ is much smaller than the length scale 
for significant variations in $\phi$.  The matrix elements can 
therefore be expanded in derivatives of $\phi$
by generalizing the methods developed in 
Refs. \cite{Braaten-Nieto:2,Andersen-Braaten:2} 
to the case of a time-dependent state. 
Including terms through first order in $g^{3/2} |\phi|$
and through first order in derivatives of $\phi$, 
we obtain \cite{Andersen-Braaten:3}
\begin{eqnarray}
\langle \tilde\psi^\dagger \tilde\psi \rangle 
	&\approx& {1 \over 24 \pi^2} g^{3/2} |\phi|^3,
\label{ave-normal}
\\
\langle \tilde\psi \tilde\psi \rangle 
	&\approx& {1 \over 8 \pi^2} g^{3/2} |\phi| \phi^2
	\;-\; {i m \over 8 \pi^2 \hbar} g^{1/2} |\phi|^{-3} \phi^2 
		(\phi^* \dot \phi + \dot \phi^* \phi),
\label{ave-anom}
\\
\langle \tilde\psi^\dagger \tilde\psi \tilde\psi \rangle 
	&\approx& 0,
\label{ave-triple}
\\
\langle \tilde\psi^{\dagger} \nabla \tilde\psi 
		- \nabla \tilde\psi^{\dagger} \tilde\psi \rangle
	&\approx& {1 \over 24 \pi^2} g^{3/2} |\phi|
	\left( \phi^* \nabla \phi - \phi \nabla \phi^* \right).
\label{ave-grad}
\end{eqnarray}
To verify the continuity equation $\dot \rho + \nabla \cdot {\bf j} = 0$
to first order in $g^{3/2}|\phi|$ and to second order in gradients of $\phi$, 
we must include terms in
$\langle \tilde\psi^\dagger \tilde\psi \rangle $ and 
$\langle \tilde\psi \tilde\psi \rangle $
that are second order in gradients of $\phi$.  
These terms are rather complicated \cite{Andersen-Braaten:2},
but fortunately they are not needed to derive the equation for the time 
evolution of $\rho$ in the Thomas-Fermi limit.
By differentiating the continuity equation  with respect to time, 
using (\ref{qfe-phi}) to eliminate $\dot \phi$ 
from the time derivative of ${\bf j}$, and 
using (\ref{rho-SC}) to eliminate $\phi$ in favor of $\rho$, 
we can obtain a closed equation for $\rho$.
Keeping only terms through first order in 
$(g^3 \rho)^{1/2}$ and through second order in gradients of $\rho$, 
this equation is
\begin{equation} 
\ddot \rho \;=\; 
{\hbar^2 g \over 4 m^2} \nabla \cdot \left\{ \rho \nabla \left[ 
\rho \;-\; {4 m \over \hbar^2 g} (\mu - V) 
\;+\; {1 \over 6 \pi^2} g^{3/2} \rho^{3/2} \right] \right\}. 
\label{rho-doubledot} 
\end{equation} 
This equation gives the time-dependence of the density in the Thomas-Fermi 
limit and to first order in the semiclassical expansion. 
The ground state number density $\rho_0$ in this approximation satisfies 
\begin{equation} 
\left[ 1 + {1 \over 6 \pi^2} (g^3 \rho_0)^{1/2} \right] \rho_0 
\;=\; {4 m \over \hbar^2 g} ( \mu - V ) .
\label{rho-bar} 
\end{equation} 
To deduce the equations for small-amplitude oscillations 
of the condensate, we substitute $\rho = \rho_0 + \tilde \rho$ 
into (\ref{rho-doubledot}) and linearize in 
$\tilde \rho$: 
\begin{equation} 
\ddot {\tilde \rho} \;=\; 
{\hbar^2 g \over 4 m^2} 
\nabla \cdot \left\{ \rho_0 \; \nabla \left[ \tilde \rho 
+ {1 \over 4 \pi^2} g^{3/2} \rho_0^{1/2} \tilde \rho \right] \right\}. 
\label{rhotilde-doubledot} 
\end{equation} 

In the mean-field approximation, 
the $(g^3 \rho_0)^{1/2}$ corrections in (\ref{rho-bar}) 
and (\ref{rhotilde-doubledot}) are neglected. 
The solution for the ground state number density $\bar \rho_0$ 
vanishes at points ${\bf r}$ where $V({\bf r}) > \mu$ 
and is given by 
$\bar \rho_0 = 4 m (\mu - V)/\hbar^2 g$ at points where 
$V({\bf r}) < \mu$. 
Assuming that $\tilde \rho$ has the time dependence 
$\exp(i\bar \omega t)$, we obtain the eigenvalue equation 
$H_0 {\tilde \rho} = \bar \omega^2 {\tilde \rho}$, where 
\begin{equation} 
H_0 \;=\; 
- {\hbar^2 g \over 4 m^2} \nabla \cdot \left( \bar \rho_0 \; \nabla \right). 
\label{H0} 
\end{equation} 
The ``hamiltonian'' $H_0$, whose eigenvalues give the mean-field 
approximations to the oscillation frequencies in the Thomas-Fermi limit, 
was first derived by Stringari \cite{Stringari}. 

To obtain the semiclassical correction to the frequencies, we expand 
(\ref{rhotilde-doubledot}) to first order in $(g^3 \bar \rho_0)^{1/2}$. 
The solution of (\ref{rho-bar}) to first order in 
$(g^3 \bar \rho_0)^{1/2}$ is 
$\rho_0 = \bar \rho_0 [1 - (g^3 \bar \rho_0)^{1/2}/6 \pi^2]$. 
Inserting this into (\ref{rhotilde-doubledot}) 
and assuming that $\tilde \rho$ has the time-dependence 
$\exp(i \omega t)$, 
we obtain an eigenvalue equation for $\omega$. 
Absorbing a factor of 
$1 + g^{3/2} {\bar \rho}_0^{1/2}/(8 \pi^2)$ into $\tilde \rho$, 
our equation becomes the eigenvalue equation for a hermitian operator: 
$(H_0 + H_1) {\tilde \rho} = \omega^2 {\tilde \rho}$, where 
\begin{equation} 
H_1 \;=\; - {\hbar^2 g^{5/2} \over 96 \pi^2 m^2} 
\left( {\bar \rho_0}^{3/2} \nabla ^2 
\;+\; \nabla ^2 {\bar \rho_0}^{3/2} \right). 
\label{H1} 
\end{equation} 
The eigenvalues of $H_0 + H_1$ give the oscillation 
frequencies in the Thomas-Fermi limit through first order in the 
semiclassical expansion. 
The semiclassical corrections to the frequencies can be obtained by 
using the standard results of time-independent perturbation theory. 
If $\bar \omega_n^2$ is a nondegenerate eigenvalue of $H_0$ with 
eigenfunction ${\tilde \rho}_n$, then the semiclassical correction 
$\delta \omega_n^2= \omega_n^2- \bar \omega_n^2$ is 
\begin{equation} 
\delta \omega_n^2 \;=\; - {\hbar^2 g^{5/2}\over 48 \pi^2 m^2} 
{ {\rm Re} \int d^3r {\bar \rho_0}^{3/2} 
{\tilde \rho_n}^* \nabla ^2 {\tilde \rho_n} 
\over \int d^3r \left| {\tilde \rho_n} \right|^2 }. 
\label{eigenvalue} 
\end{equation} 
If the eigenvalue $\bar \omega_n^2$ is degenerate, then the expression 
(\ref{eigenvalue}) holds only in the basis consisting of states 
that diagonalize the restriction 
of $H_1$ to the subspace of states with eigenvalue $\bar \omega_n^2$. 
Note that the correction (\ref{eigenvalue}) is positive or zero. 
This follows from the fact that $H_1$ is proportional to the 
anticommutator of the positive operator ${\bar \rho_0}^{3/2}$ 
and the positive-semidefinite operator $- \nabla^2$. 
The correction (\ref{eigenvalue}) is zero only if the laplacian 
of the eigenfunction ${\tilde \rho}_n$ vanishes.

We first consider a condensate trapped in the 
spherically symmetric parabolic potential 
$V(r) = m \omega_0^2 r^2/2$. 
The ground state number density ${\bar \rho}_0$ 
in the mean-field approximation and in the Thomas-Fermi limit 
vanishes outside a sphere of radius $R$ given by 
$N = m^2 \omega_0^2 R^5/(15 \hbar^2 a)$. Inside that sphere, 
it has the form 
\begin{equation} 
{\bar \rho}_0(r) \;=\; {2 m^2 \omega_0^2 \over \hbar^2 g} (R^2 -r^2), 
\qquad r < R. 
\end{equation} 
The normal modes for oscillation of the condensate around the 
ground state can be labelled by a radial quantum number $n$ 
and angular momentum quantum numbers $\ell$ and $m$. 
The mean-field frequencies $\bar \omega_{n \ell}$ depend 
on the quantum numbers $n$ and $\ell$, 
and were first determined by Stringari \cite{Stringari}: 
\begin{equation} 
\bar \omega_{n \ell}^2 \;=\; 
\left[ 2 n^2 + 3 n + (2 n + 1) \ell \right] \omega_0^2. 
\end{equation} 
The $n=0$ modes are surface modes and their eigenfunctions are 
$r^\ell Y_{\ell m}(\theta,\phi)$. Since they have a vanishing laplacian, 
the semiclassical correction to the 
frequencies of these surface modes vanishes: $\delta \omega_{0 \ell}^2 = 0$. 
The eigenfunctions for $n=1$ are 
$( R^2 - {2 \ell + 5 \over 2 \ell + 3} r^2) r^\ell Y_{\ell m}(\theta,\phi)$ 
for $r<R$. Inserting this into (\ref{eigenvalue}),
we find that the shifts in the frequencies are 
\begin{equation} 
\delta \omega_{1 \ell}^2 \;=\; 
{3 (2 \ell + 3) ({1 \over 2})_{\ell+4} \over \sqrt{2}(\ell+4)!} 
\left( {m \omega_0 a R \over \hbar} \right) \omega_0^2,
\label{del-1l} 
\end{equation} 
where $(z)_n = \Gamma(z+n)/\Gamma(z)$ is the Pochhammer symbol. 
The fractional shift in the frequency is proportional to 
$m \omega_0 a R / \hbar$, which scales like $N^{1/5} a^{6/5} \omega_0^{3/5}$. 
It therefore increases very slowly with the number of atoms 
in the condensate.

We next consider a condensate trapped in the axially 
symmetric parabolic potential 
$V(\rho,z) = m ( \omega_\perp^2 \rho^2 
+ \omega_\parallel^2 z^2)/2$. 
The ground state number density 
in the mean-field approximation and in the Thomas-Fermi limit 
takes the form 
\begin{equation} 
{\bar \rho}_0({\bf r}) \;=\; 
{2 m^2 \omega_\perp^2 \over \hbar^2 g} (R^2-\rho^2-\lambda^2 z^2), 
\qquad \rho^2+\lambda^2 z^2 < R^2, 
\end{equation} 
where $\lambda =\omega_\parallel/\omega_\perp$ and $R$ 
is given by $N = m^2 \omega_\perp^2 R^5/(15 \hbar^2 a \lambda)$. 
The eigenfunctions have the form 
$\rho^{|m|} e^{im\phi} P(z,\rho^2)$, 
where $m$ is the azimuthal angular momentum 
quantum number and $P$ is a polynomial in $z$ and $\rho^2$. 
The number of independent solutions for which $P$ is a 
$k$'th order polynomial in $z$ is ${\rm int}(1+k/2)$. 
For $k=0$ and $k=1$, the frequencies 
in the mean-field approximation are 
$\bar \omega_{0m}^2=|m|\omega_\perp^2$ and 
$\bar \omega_{1m}^2=(|m| + \lambda^2) \omega_\perp^2$, respectively. 
The corresponding eigenfunctions are 
$\rho^{|m|} e^{im\phi}$ and $z \rho^{|m|} e^{im\phi}$. 
Since these eigenfunctions have vanishing laplacian, 
the semiclassical corrections to the frequencies vanish: 
$\delta\omega_{0m}^2 = \delta\omega_{1m}^2 =0$. 
For each value of $m$, there are two independent eigenfunctions with $k=2$. 
Their frequencies in the mean-field approximation are 
\begin{equation} 
\bar\omega_{2m\pm}^2 \;=\; 
\left( 2(|m|+1) + \mbox{${3\over2}$} \lambda^2 
\pm \sqrt{(|m|+2)^2 - (|m|+4) \lambda^2 + \mbox{${9\over4}$} \lambda^4} 
\right) \omega_\perp^2. 
\label{omega-MF} 
\end{equation} 
These were first derived by Stringari for $m=0$ \cite{Stringari} 
and by Ohberg et al. and by Fliesser et al. for general $m$ 
\cite{Ohberg-Fliesser}. The eigenfunctions have the form 
$( R^2 + \alpha_{m \pm} \rho^2 + \beta_{m \pm} \lambda^2 z^2 ) 
\rho^{|m|} e^{im\phi}$, 
where $\alpha_{m \pm}$ and $\beta_{m \pm}$ are complicated functions of 
$|m|$ and $\lambda$. Using (\ref{eigenvalue}), we find that the 
semiclassical corrections to the frequencies are 
\begin{eqnarray} 
\delta\omega_{2m\pm}^2 &=& 
{ 3 ({1\over2})_{|m|+4} f_\pm(|m|,\lambda) \over \sqrt{2} (|m|+4)! } 
\left( {m \omega_\perp a R \over \hbar} \right) \omega_\perp^2 , 
\nonumber 
\\ 
f_\pm(|m|,\lambda) &=& 
|m|+1 + \mbox{${1\over2}$}\lambda^2 \;\pm\; 
{ (|m|+1)(|m|+2) - {1\over2} \lambda^2 + {3\over4} \lambda^4 
\over \sqrt{(|m|+2)^2 - (|m|+4) \lambda^2 + {9\over4} \lambda^4} }. 
\label{omega-shift} 
\end{eqnarray} 
If we set $\lambda=1$, this reduces to (\ref{del-1l}) with $\ell \to |m|$.

The most accurate measurement to date of an oscillation frequency for 
a Bose-Einstein condensate was carried out by 
Stamper-Kurn et al. \cite{Stamper-Kurn}. They trapped $N=1.5 \times 10^{7}$ 
atoms of $^{23}$Na, whose scattering length is $a=2.75$ nm, 
in an axially symmetric potential. 
The axial trapping frequency was measured to be 
$\omega_\parallel/2 \pi = 16.93(2)$ Hz and the transverse trapping 
frequency was estimated to be $\omega_\perp/2 \pi = 230$ Hz, 
so that $\lambda \approx 0.074$. They measured the ratio 
$\omega_{20-}/\omega_\parallel$ of the lowest quadrupole 
oscillation frequency to the axial trap frequency to be 1.569(4). 
The fractional uncertainty in the measurement is 0.25\%. 
The mean-field prediction from (\ref{omega-MF}) is 
$\bar \omega_{20-}/\omega_\parallel=1.5806$, 
which is larger by about 3 standard deviations. 
Including the semiclassical correction (\ref{omega-shift}), 
we obtain the prediction 1.5813. 
The fractional increase is 0.04\%, which is about ${1 \over 6}$ 
of a standard deviation. Thus there is a several standard 
deviation discrepancy between the prediction and the measurement.

In order to assess the significance of the discrepancy, we must consider other 
sources of corrections to the frequency, including edge effects, 
nonlinear effects, and nonzero temperature effects. 
The correction from edge effects arises from the breakdown of the
Thomas-Fermi approximation near the edge of the condensate. 
The fractional correction scales like $(\hbar/m \omega R^2)^2$ 
\cite{Fetter-Feder}, which is approximately 4\% 
for the experiment in Ref. \cite{Stamper-Kurn},
so these corrections are probably not negligible.
This source of theoretical error can be eliminated by avoiding 
the Thomas-Fermi limit, and instead calculating the frequencies numerically 
by solving the mean-field equations. 
The frequency shift from nonlinear effects has been studied by 
Dalfovo et al. \cite{Dalfovo-Minniti-Pitaevskii}. 
The fractional shift for the lowest quadrupole mode was found to be 
$+0.09 A^2$, where $A$ is the relative magnitude of the oscillation 
in the length of the condensate. Taking the value $A \approx 0.1$ from 
Ref. \cite{Stamper-Kurn}, we find that the fractional frequency shift is 
0.9\%. This source of error can be eliminated experimentally by 
decreasing the amplitude of the excitation of the condensate. 
It was suggested in Ref. \cite{Stamper-Kurn} that the discrepancy 
in the frequency can be explained by nonzero temperature corrections, which 
have the correct sign. While there have been studies 
of the temperature dependence of the frequencies \cite{review}, 
there don't seem to be any quantitative treatments 
for the low temperatures 
appropriate to the experiment of Ref. \cite{Stamper-Kurn}.

We have calculated the semiclassical corrections to the frequencies 
for small oscillations of a Bose-Einstein condensate in 
the Thomas-Fermi limit. The frequency shifts are less than an order 
of magnitude smaller than the uncertainties in 
some previous frequency measurements. 
These shifts should therefore be measurable by more accurate experiments. 
Accurate theoretical predictions require going beyond the 
Thomas-Fermi limit in calculating the semiclassical frequency shifts. 
Separating the semiclassical correction from that due to nonzero 
temperature may require the development of new techniques 
for measuring the temperatures of extremely cold condensates. 
A measurement of the semiclassical frequency shift in agreement 
with theoretical predictions would mark a new threshold in our 
quantitative understanding of Bose-Einstein condensed gases. 

While this work was being prepared for publication, 
a preprint by Pitaevskii and Stringari \cite{P-S} appeared 
in which the authors also 
calculate the semiclassical corrections to the oscillation frequencies 
of a trapped Bose-Einstein condensate. 
Their results are in perfect agreement with ours.

We thank J.O. Andersen for valuable discussions. 
This work was supported in part by the U.S.
Department of Energy Division of High Energy Physics
and by a Research Experience for Undergraduates site grant
at the Ohio State University.

%%%%%%%%%%%%%%%%%%%%%%%%%%%%%%  REFERENCES  %%%%%%%%%%%%%%%%%%%%%%%%%%%%%%

\end{document}